\documentclass[twocolumn]{aastex63}
 
\usepackage{graphicx}
\usepackage{subfigure}
\usepackage{multirow}
\usepackage{comment}
\usepackage{natbib}
\usepackage{hyperref}
\usepackage{mathtools}
\usepackage{mathrsfs}
\usepackage{fontenc}
\usepackage{color}
\usepackage{url}
\usepackage{hyperref}
\usepackage{pifont}
\usepackage{textcomp}
\usepackage{color}
\usepackage{multirow}
\usepackage{booktabs}
\usepackage{array}

\def\hi{H~{\sc i}~} 
\def\h2{H$_2$} 
\def\nhi{$N${\sc (H~i)}}

\def\zabs{$z_{\rm abs}$} 
 
\def\zemi{$z_{\rm em}$~} 
\def\lya{Ly$\alpha$~} 
\def\llya{$L_{\rm Ly\alpha}$~}

\def\kms{km s$^{-1}$} 
\newcommand{\qso}{SDSS~J$011852+040644$}

\begin{document}

\title{Discovery of a damped \lya galaxy at $z \sim 3$ towards the quasar \qso}
\shorttitle{A new $z \sim 3$ damped \lya galaxy}
\shortauthors{Joshi et al}

\author{Ravi Joshi}
\affil{Kavli Institute for Astronomy and Astrophysics, Peking University, Beijing 100871, China}

\author{Michele Fumagalli}
\affil{Dipartimento di Fisica G. Occhialini, Universit\`a degli Studi di Milano Bicocca, Piazza della Scienza 3, 20126 Milano, Italy}
\affil{Institute for Computational Cosmology, Durham University, South Road, Durham, DH1 3LE, UK}
\affil{Centre for Extragalactic Astronomy, Durham University, South Road, Durham, DH1 3LE, UK}

\author{Raghunathan Srianand}
\affil{Inter-University Centre for Astronomy and Astrophysics, Post Bag 4, Ganeshkhind, Pune 411007, India }

\author{Pasquier Noterdaeme}
\affil{Sorbonne Universit\`e, CNRS, UMR 7095, Institut d'Astrophysique de Paris, 98 bis bd Arago, 75014 Paris, France }

\author{Patrick  Petitjean}
\affil{Sorbonne Universit\`e, CNRS, UMR 7095, Institut d'Astrophysique de Paris, 98 bis bd Arago, 75014 Paris, France } 

\author{Marc Rafelski}
\affil{Space Telescope Science Institute, Baltimore, MD 21218, USA}
\affil{Department of Physics \& Astronomy, Johns Hopkins University, Baltimore, MD 21218, USA}

\author{Ruari Mackenzie}
\affil{Centre for Extragalactic Astronomy, Durham University, South Road, Durham, DH1 3LE, UK}
\affil{Department of Physics, ETH Zurich, Wolfgang-Pauli-Strasse 27, 8093 Zurich, Switzerland }

\author{Qiong Li}
\affil{Kavli Institute for Astronomy and Astrophysics, Peking University, Beijing 100871, China}
\affil{Department of Astronomy, School of Physics, Peking University, Beijing 100871, China}

\author{Zheng Cai}
\affil{Department of Astronomy, Tsinghua University, Beijing 100084, China}

\author{D. Christopher Martin}
\affil{Cahill Center for Astrophysics, California Institute of Technology, 1216 East California Boulevard, Mail Code 278-17, Pasadena, \\ California 91125, USA}

\author{Siwei Zou}
\affil{Kavli Institute for Astronomy and Astrophysics, Peking University, Beijing 100871, China}

\author{Xue-Bing Wu}
\affil{Kavli Institute for Astronomy and Astrophysics, Peking University, Beijing 100871, China}
\affil{Department of Astronomy, School of Physics, Peking University, Beijing 100871, China}

\author{Linhua Jiang}
\affil{Kavli Institute for Astronomy and Astrophysics, Peking University, Beijing 100871, China}

\author{Luis C.  Ho}
\affil{Kavli Institute for Astronomy and Astrophysics, Peking University, Beijing 100871, China}
\affil{Department of Astronomy, School of Physics, Peking University, Beijing 100871, China}

\correspondingauthor{Ravi Joshi}
\email{rvjoshirv@gmail.com}

\begin{abstract}
We report the detection of the host galaxy of a damped \lya system
(DLA) with log \nhi\ $ [\rm cm^{-2}]$ = $21.0 \pm 0.10$ at $z \approx
3.0091$ towards the background quasar \qso\ using the Palomar Cosmic
Web Imager (PCWI) at the Hale (P200) telescope. We detect \lya
emission in the dark core of the DLA trough at a 3.3$\sigma$
confidence level, with \lya luminosity of \llya $\rm = (3.8 \pm 0.8)
\times 10^{42}\ erg\ s^{-1}$, corresponding to a star formation rate
of $\gtrsim 2\ \rm M_{\odot}\ yr^{-1}$ (considering a lower limit on Ly$\alpha$ escape fraction 
$f_{esc}^{Ly{\alpha}} \sim 2\%$) as typical for Lyman break galaxies
at these redshifts. The \lya emission is blueshifted with respect to
the systemic redshift derived from metal absorption lines by $281 \pm
43$~\kms. The associated galaxy is at very small impact parameter of
$\lesssim 12 \rm\ kpc$ from the background quasar, which is in line
with the observed anticorrelation between column density and impact
parameter in spectroscopic searches tracing the large-scale
environments of DLA host galaxies.
 
\end{abstract}


\section{Introduction}
\label{sec:intro}
The evolution of galaxies is significantly influenced by the physical state of gas in and around the central star-forming regions. Observations of local galaxies indicate that the atomic and molecular hydrogen, which make up most of the mass in the interstellar medium (ISM) of galaxies, closely trace the star-formation rate and are  the key elements that participate in inflows and outflows
\citep{Bigiel2008AJ....136.2846B,Genzel2010MNRAS.407.2091G,Cortese2011MNRAS.415.1797C,Janowiecki2017MNRAS.466.4795J}.

Unfortunately, mapping neutral hydrogen (H{\sc~i}) gas in emission
from galaxies is difficult at even moderate redshifts
\citep{Kanekar2016ApJ...818L..28K}.  This low density
  gas imprints, however, absorption lines on the spectra of an unrelated bright
  background sources, which offers a  powerful tool to study the physical
  and chemical properties of the intervening gas in a luminosity independent
  manner \citep[see][for a review]{Wolfe2005ARA&A..43..861W}. At high
redshift, most of our knowledge of \hi\ gas primarily comes from a particular class of absorption line systems, the
damped \lya absorbers (DLAs) seen in quasar spectra. With an \hi
column density of $\ge 2 \times 10^{20}~\rm cm^{-2}$, DLAs account for the
bulk ($> 80\%$) of the neutral hydrogen in the early Universe
\citep{Peroux2003MNRAS.346.1103P,Prochaska2009ApJ...696.1543P,Noterdaeme2009A&A...505.1087N,Noterdaeme2012A&A...547L...1N}.
  
  Moreover, DLAs appear to be linked to star-forming regions, as
  evidenced by the metallicity evolution of DLAs with redshift
  \citep{Rafelski2012ApJ...755...89R,Rafelski2014ApJ...782L..29R,Jorgenson2013MNRAS.435..482J}
  and the velocity spread of low-ion absorption lines
  \citep[see][]{Wolfe2005ARA&A..43..861W}. The average properties of
  \lya emission from DLAs, inferred from the stacking experiment of
  hundreds of DLAs from the Sloan Digital Sky Survey (SDSS), further
  indicate a connection between star formation activity and outflows
  in DLA host galaxies \citep[see
    also,][]{Rahmani2010MNRAS.409L..59R,Noterdaeme2014A&A...566A..24N,Joshi2017MNRAS.465..701J}.
  Therefore, establishing a direct association of the \hi gas seen in
  absorption with emission from galaxies is a useful way to probe the
  link between the \hi gas and star formation at high redshift.

Earlier efforts to detect DLA host galaxies either in continuum emission or nebular line emission have been moderately successful in completely blind surveys, with a detection rate of $\sim 10\%$ \citep{Moller2004A&A...422L..33M}, with several studies mostly resulting in non-detections
\citep{Kulkarni2000ApJ...536...36K,Christensen2009A&A...505.1007C,Fumagalli2015MNRAS.446.3178F}.
Leveraging the observed correlation between luminosity and metallicity in galaxies
\citep{Tremonti2004ApJ...613..898T,Ledoux2006A&A...457...71L,Moller2013MNRAS.430.2680M,Christensen2014MNRAS.445..225C}, recent campaigns have focused instead on metal-rich DLAs, resulting in a far higher detection rate of $\approx 65\%$ \citep{Fynbo2010MNRAS.408.2128F,Fynbo2011MNRAS.413.2481F,Krogager2017MNRAS.469.2959K,Ranjan2020A&A...633A.125R}.

In spite of these numerous attempts, however, only $\approx20$ DLAs at redshift $\gtrsim 2$
have been associated directly to counterparts in emission
\citep[see][]{Krogager2017MNRAS.469.2959K}. This low detection rate
could be attributed either to the faint nature of DLA galaxies (which
become difficult to image at close separation from bright background
quasars), or to their dusty nature, or to high \hi column density, or
yet again to the fact that only a fraction of the DLA population is directly
connected to active star-formation.

More recently, interferometers such as the Atacama Large
Millimeter/submillimeter Array (ALMA) have overcome the dust bias,
with the detection of $\sim$ 10 molecular gas-rich systems using CO
rotational transitions and the atomic [CII] line
\citep{Neeleman2016ApJ...820L..39N,Neeleman2018ApJ...856L..12N,Fynbo2018MNRAS.479.2126F,Klitsch2019MNRAS.482L..65K}.
       So far, these studies have focused on tracing relatively
        high-metallicity systems, finding the DLA hosts at relatively
        large impact parameters, $\sim 16 - 45 \rm \ kpc$, and with high
        molecular gas masses of $10^{10}-10^{11} M_{\odot}$. Following these successes, 
        efforts to detect more representative DLAs are ongoing.

Furthermore, the use of integral field spectrographs (IFSs) at 8-10m class telescopes has proven to be a very efficient tool for searching DLA galaxies and for characterizing their environment out to several hundreds of kiloparsecs
\citep{Peroux2011MNRAS.410.2251P,Peroux2012MNRAS.419.3060P,Fumagalli2017MNRAS.471.3686F,Mackenzie2019MNRAS.tmp.1435M}.
For example, using the MUSE IFS at the VLT telescope, \citet{Fumagalli2017MNRAS.471.3686F} have detected a tantalizing
example of extended \lya emission tracing gas in a region of about 50 kpc near a $z\approx 3$ DLA. This region
hosts multiple galaxies, possibly in a filament, 
with \lya emission induced by {\it in-situ} star formation likely triggered by interactions. 

Moreover, in a recent MUSE
survey of 6 DLAs at $z \sim 3$,
\citet{Mackenzie2019MNRAS.tmp.1435M} have obtained a high detection rate of galaxies up to $\approx 80\%$ within 1000~\kms\ of the DLAs, with impact
parameters ranging between 25 and 280 kpc. 
Notably, in contrast to previous searches, the blind survey of \citet{Mackenzie2019MNRAS.tmp.1435M} has yielded detections of multiple galaxies also for low metallicity systems (see their figure 9), including a galaxy group associated with a metal poor DLA ($Z/Z_{\odot} \approx - 2.33$). With a small but representative sample, using cosmological simulations, these authors have 
been able to place constraints on the typical mass of halos that host DLAs in the range $10^{11}-10^{12}
M_{\odot}$.

To further understand the link between \hi gas and star formation near
the peak of the cosmic star-formation activity, we have started a
survey to search for high-redshift ($z \gtrsim 3$) DLA host galaxy
counterparts in Ly$\alpha$ emission using the Palomar Cosmic Web
Imager (PCWI). In this article, we present results from a pilot
  observation that traces the large scale environment of a strong
  intervening DLA with $\log$~\nhi\ $\rm [cm^{-2}] = 21.0\pm0.10~$ at
  \zabs$\approx 3.0091$ out to 80 kpc. Our observations lead to the
  discovery of the host galaxy, revealing a direct association of the
  absorbing gas with star-formation, with no other counterpart within
  the field of view   $20\arcsec \times
   40\arcsec$. This paper is organized as follows. Section 2
describes the sample selection. In Section 3, we present the
observations and data reduction. In Section 4, we present results of
our analysis, followed by a discussion and conclusion in Section
5. Throughout, we have assumed a flat Universe with $H_0 =$
70~\kms\ $\rm Mpc^{-1}$, $\Omega_{\rm m} = 0.3$ and $\Omega_{\rm
  \Lambda} = 0.7$.

\begin{figure}
\centering
\includegraphics[clip,trim={0.8cm 2.8cm 0.1cm 2.3cm},height=12.0cm,width=8.6cm]{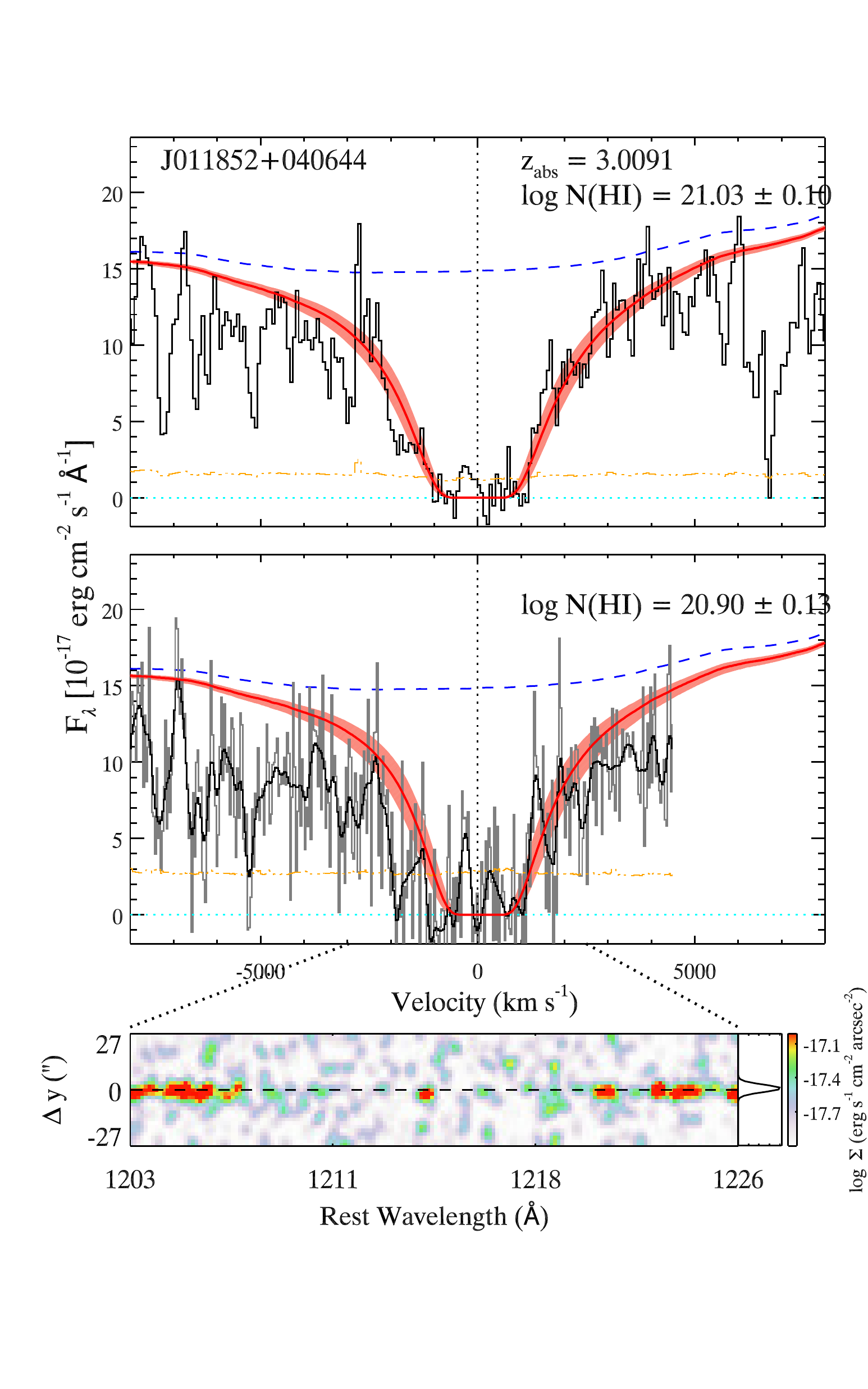}
  \caption {\emph{Top panel:} The \lya\ absorption profile in the SDSS
    spectrum (black histogram) in the velocity scale with respect to
    \zabs  $\sim$ 3.0091. The estimated unabsorbed quasar continuum is
    shown as blue dashed curve along with the error spectrum using the
    dot-dashed curve. The continuum template modified by the damped
    \lya absorption is shown with a red solid line, marking the
     profile uncertainty  corresponding to 1$\sigma$ error in column density with a red shaded region.
    \emph{Middle panel:} 1D quasar spectrum from the original PCWI
    data (gray histogram) and following resampling at the SDSS
    resolution of 2.5\AA (black histogram). A new model fit derived on
    PCWI data, which is consistent with the SDSS estimate, is also
    shown. \emph{Bottom panel:} 2D quasar spectrum constructed from
    PCWI data cube. The trace of the quasar is shown as dashed line.}
  \label{fig:DLACORE} 
    \end{figure}

\begin{figure}
\centering
\includegraphics[clip,trim={0 8.6cm 0.5cm 0.5cm},height=5.5cm,width=8.5cm]{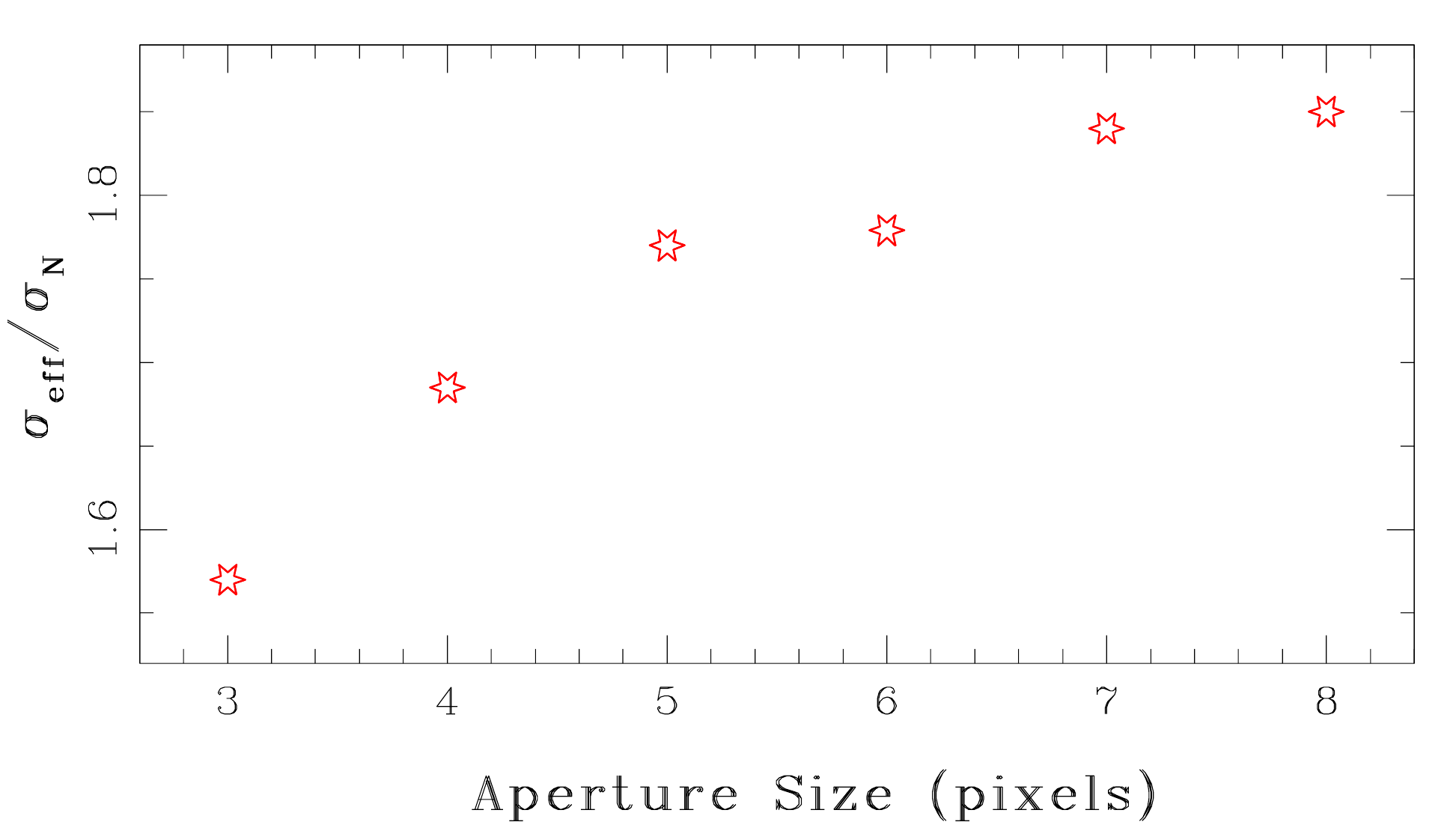}
  \caption{ Ratio between the flux dispersion in apertures of varying size ($\sigma_{\rm eff}$) and the error computed propagating the pixel standard deviation ($\sigma_{\rm N}$), which is useful to assess the impact of correlated noise within the PCWI data cube.}
  \label{fig:noise} 
    \end{figure}

\section{Sample  selection}
\label{sec:sample}
Using the compilation of thousands of DLAs from SDSS
\citep{Noterdaeme2012A&A...547L...1N}, we have selected a subset
having high H{\sc~i} column density, with log \nhi $\rm [cm^{-2}]$
$\ge 21~\rm $, which provides the favorable environment for star
formation and thus is likely to trace regions in close proximity to
star-forming galaxies
\citep{Krogager2012MNRAS.424L...1K,Altay2013MNRAS.434..748A}.
Moreover, at these high column densities, the Ly$\alpha$ absorption
has a dark  (optically thick) core, which spreads
over at least 7 times the average full-width at half-maximum (FWHM,
$\approx$ 160 $ \rm km~ s^{-1})$ of the instrumental profile of the
SDSS spectrograph. This makes it possible to search for \lya\ emission
lines within the spectrum.

We
consider only the DLAs detected in SDSS spectra with a median
continuum-to-noise ratio $> 3$ which ensure an accurate determination
of the H{\sc~i} column density. In addition, we avoid DLAs that are proximate to the quasars by considering only systems with velocity offsets of $> 5000\ \rm
km\ s^{-1}$ with respect to the quasar emission redshift.
We also exclude  sightlines showing broad absorption lines from quasar outflows. To avoid introducing a metallicity bias, we do not preselect targets based on metal
lines (e.g., Si{\sc~ii}, Fe{\sc~ii}, C{\sc~ii}).

In order to maximise the detection rate of Ly$\alpha$ emission, we further search the spectra to identify systems with tentative Ly$\alpha$ emission
(non-zero flux) within the absorption trough where the quasar continuum goes to zero (i.e., the dark core). 
For this, we  avoid the regions with bright sky emission  to exclude the false positives
due to residuals of the sky subtraction. 
Due to the finite fiber size of SDSS, this step introduces a selection effect, that is detections are expected to  primarily occur at small impact parameters of $\lesssim 15~\rm kpc$ (see below). 
An example of system selected in this way, which is also the target of our pilot observations,  is
shown in the upper panel of Figure~\ref{fig:DLACORE}. 

 Following visual inspection to remove systems with clear sky residuals
or artifacts, this selection resulted in a unique set of 13 DLAs (out
of 608) with absorption redshifts $z\ge 2.9$.  Among them, 10 systems
lie at declinations that Hale (P200) can reach and are suitable for
P200 observations. As a further verification of the presence of
possible \lya\ emission, we have also examined the multi-epoch
observations from SDSS, which exists for 3 DLAs in our
sample. Reassuringly, all 3 systems shows Ly$\alpha$ emission in
spectra at different epochs, albeit with low signal-to-noise. In order
to be able to detect the minimum \lya emission flux of $\sim 2 \times
10^{-17} \rm erg\ s^{-1}\ cm^{-2}$ found in our sample, an integration
of $\sim 1$hr on source and $\sim 1$hr on sky would allow us to detect
the emission feature at more than $\sim 5 \sigma$ level. In what
follows, we present the results for the first target successfully
observed so far.

\section{Observations and Data reduction}
\label{sec:obs}

 We have performed observations of the quasar \qso\ (\zemi $\approx
 3.226$) with an intervening DLA system from our selection above
 (\zabs\ $\approx 3.0091$ and log \nhi $\rm [cm^{-2}] =21.0\pm0.1$)
 using the Palomar Cosmic Web Imager (PCWI) instrument mounted on the Hale 5 meter telescope on
 Mt. Palomar. PCWI uses a $40{\arcsec} \times 60{\arcsec}$ reflective
 image slicer with 24 slices of dimension $40{\arcsec} \times
 2.5{\arcsec}$ each.  The observations have been conducted on the
   night UT 20180816 with a clear sky and with airmass 
   ranging between 1 to 2. We have used the Richardson (MedRez)
 gratings with a slit-limited spectral resolution of $\Delta \lambda
 \sim 1$\AA.
 
 The individual exposures were acquired using the standard PCWI
 nod-and-shuffle technique, where the central 1/3 of the CCD is used
 for recording the spectrum while masking the outer 2/3 of the CCD,
 restricting the spectral bandpass to $\sim$
 150~\AA\ \citep[see,][]{Martin2014ApJ...786..106M}. Note that our DLA
 sample is preselected based on the likely presence of Ly$\alpha$
 emission within the 2 or 3 arcsec SDSS-III or SDSS-II fibre spectra.
 Thus, it is expected that the DLA host galaxy lies at small impact
 parameters (i.e, $\sim$ 8 to 12 kpc).  However, to trace the
   large scale environment around the quasar, while performing
   nod-and-shuffle, we offset the frame by 25~arcsec so that the
   quasar remains within the frame at all time. We acquired a series
 of 1200s exposures, totalling 1.6~h with the quasar at the center,
 and 1.6~h after the offset. Combined, this technique doubles the
 total integration time for the object, to 3.3~h.  This strategy
   resulted in an effective field of view of $20\arcsec \times
   40\arcsec$ with the quasar at the center which allow us to search
   for the DLA host galaxy and trace its large scale environment out
   to $\approx 80~\rm kpc$ ($\approx 10$~arcsec at $z \sim 3$).

The data are reduced with the standard PCWI pipeline
\citep{Martin2014ApJ...786..106M}. The flux calibration was performed
using the standard star BD$+$28D4211,  observed on the same
  night, and the final datacube is combined by weighting individual
exposures according to their inverse variance. In addition, the
wavelengths are converted into their values in vacuum.  The final
  cube has a pixel size of $\sim 1.5$ arcsec in the spatial direction,
  and 0.55~\AA\ in the wavelength direction.  In addition, the spatial
  resolution of PCWI cube is seeing limited along the slices in the
  short direction, which at the time of observations was $ 1.4$
  arcsec, and slit limited ($\sim$ 2.5 arcsec) in the long direction.

\begin{figure}
\centering
\includegraphics[clip,trim=30pt 0pt 140pt 100pt,height=10.5cm,width=8.5cm]{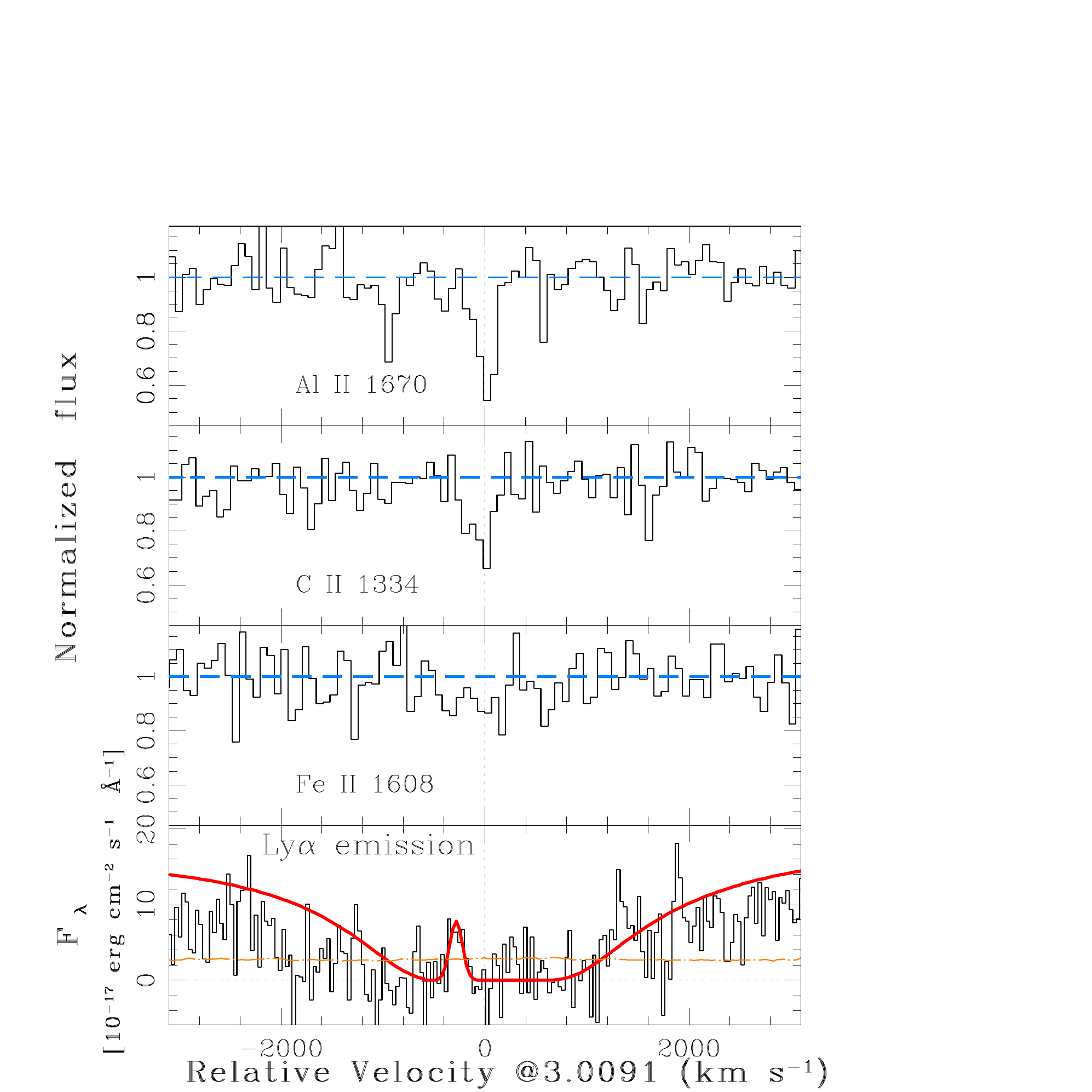}
  \caption {Velocity plots of absorption lines and the \lya emission
    from DLA host galaxy. The zero velocity is defined with respect to
    the DLA redshift at \zabs = $3.0091 \pm 0.0001$. The bottom panel shows the \lya emission 
    extracted in an aperture radius of 2 arcsec together with the best fitting
    Gaussian model. In the top panels, we plot absorption line profiles of
    Al{\sc~ii}, C{\sc~ii}, and Fe{\sc~ii} detected in SDSS spectrum.}
  \label{fig:dlaabs} 
    \end{figure}

\section{Results}

With the final PCWI datacube in hand, we first extract the quasar
spectrum in an aperture with radius of 3 arcsec which is twice the
FWHM and encompasses the total quasar emission. In
Fig.~\ref{fig:DLACORE}, we compare the DLA absorption profile in the
flux-calibrated 1D quasar spectrum from our moderate-resolution ($R
\sim 5000$) PCWI data (middle panel) with the lower-resolution ($R\sim
2000$) SDSS fiber spectrum (top panel). The best fit continuum is
  shown with a dashed (blue) curve, which is modelled with a quasar
  composite template from \citet{Harris2016AJ....151..155H} by
  adjusting the continuum power-law slope and normalisation to fit the
  quasar continuum over the \lya forest. It is clear that the DLA
absorption profile and quasar continuum level agree well in both
spectra. Next, we measure the \hi\ column densities by fitting Voigt
profiles to the \lya lines in the flux-calibrated SDSS and PCWI
spectra, as shown in Fig.\ref{fig:DLACORE}.  The modelled absorption profile touches the unabsorbed region of the
  spectrum, but also due to the low resolution, there are regions that are
  absorbed by the forest. The derived \nhi\ from the lower-resolution
SDSS spectrum is log\nhi$\rm[cm^{-2}] = 21.03 \pm 0.10$, while our
PCWI data result in log\nhi$ \rm[cm^{-2}] = 20.90 \pm 0.13$. Both
measurements are consistent with each other and with the measurement
of log\nhi $ \rm[cm^{-2}] = 21.0 \pm 0.10$ presented in
\citet{Noterdaeme2012A&A...547L...1N}.

\begin{figure*}
\centering
\includegraphics[clip,trim=5pt 0pt 110pt 2pt,height=8.2cm,width=15.4cm]{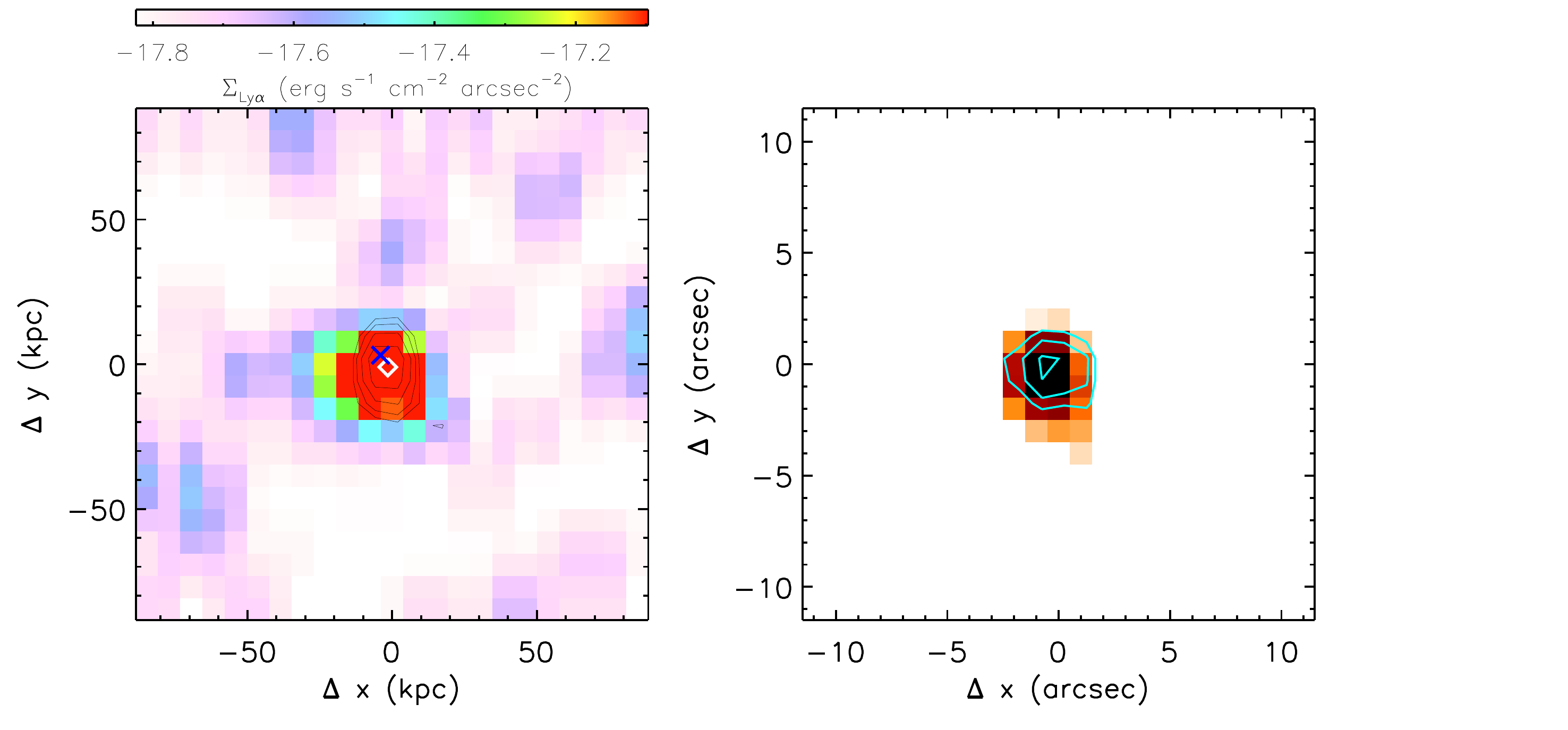}
  \caption { \emph{Left Panel:} The \lya emission map integrated over
    a velocity window from $v = -450$ \kms\ to $v = -187$ \kms,
    comprising the \lya emission feature, reveal the presence of DLA
    host galaxy at an impact parameter of $\lesssim 12$~kpc. The black
    contours mark the continuum detected source (i.e. the quasar). The
    quasar and the DLA host galaxy centers are marked with a diamond
    and a cross, respectively. \emph{Right Panel:} The white-light
    image reconstructed from the PCWI data cube showing the quasar at
    the center. The cyan contours mark the \lya emission map at the
    flux levels of 0.35, 0.45, 0.55 $\times 10^{-17} \rm erg\ s^{-1}
    cm^{-2}$.}
  \label{fig:dlamap} 
    \end{figure*}

\begin{figure}
\centering
\includegraphics[clip,trim= 0cm 0.0cm 1.0cm 1cm,height=8.2cm,width=8.2cm]{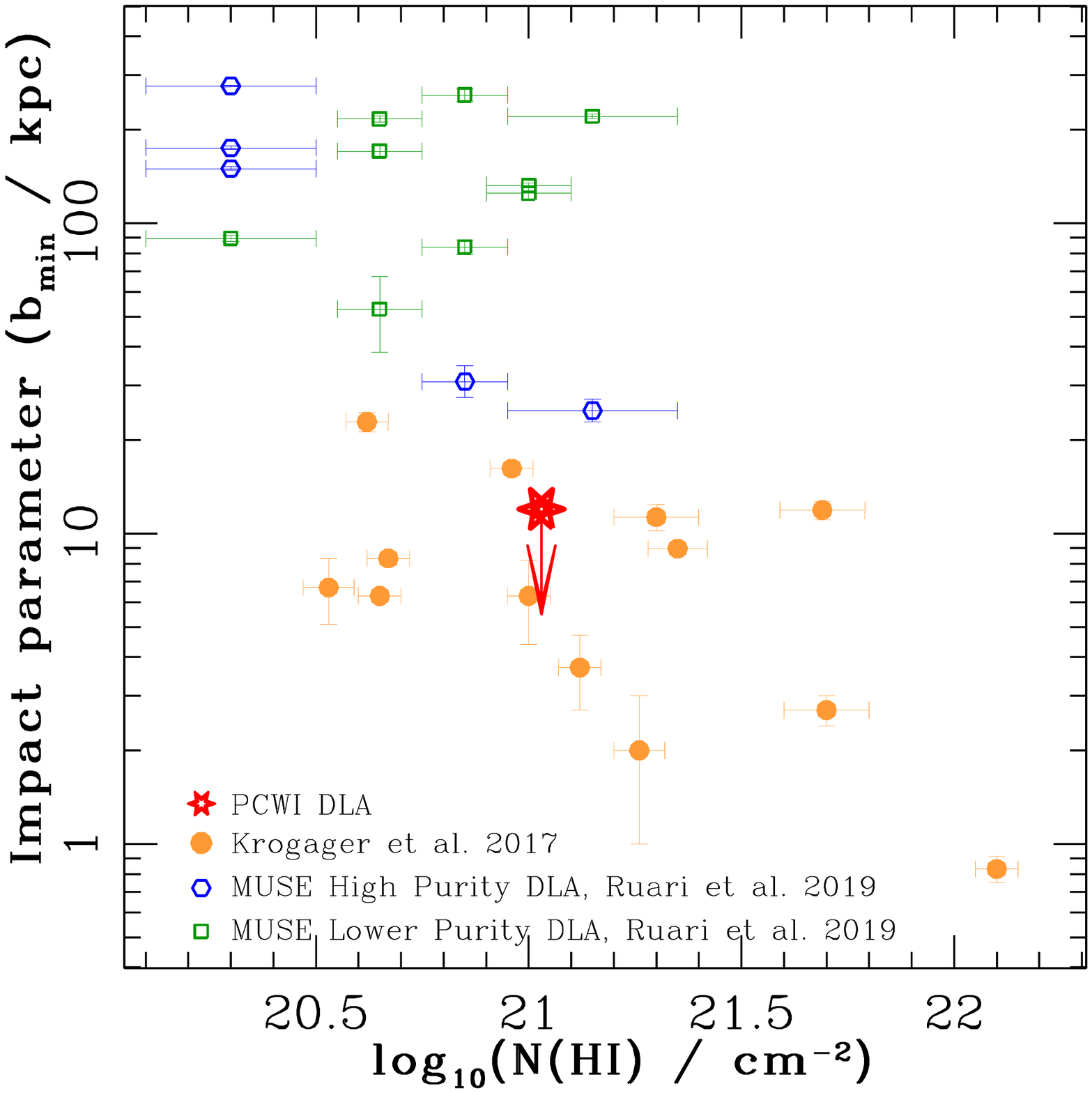}
  \caption { \hi Column density versus impact parameter relation for
    DLA host galaxies at $z > 1.5$. The filled circles show the DLAs
    identified over 25 year in long-slit based spectroscopic searches,
    compiled by \citet{Krogager2017MNRAS.469.2959K} as well as 3
    extremely strong (log\nhi\ $ \rm[cm^{-2}] \ge 21.7$)
    DLAs from \citet{Ranjan2020A&A...633A.125R}. The open diamonds show the
    galaxy population near DLAs from the recent MUSE survey by
    \citet{Mackenzie2019MNRAS.tmp.1435M}. The red star shows our PCWI
    DLA host galaxy detected at an impact parameter of $\lesssim
    12$~kpc.}
  \label{fig:nhivsb} 
    \end{figure}

In line with the pre-selection from SDSS, we clearly see an emission line signature in the DLA core at the expected position based on the SDSS spectrum. This is also evident from the reconstructed 2D spectrum from the PCWI datacube (see lower panel of Fig~\ref{fig:DLACORE}). 
To quantifying the detection significance of the \lya emission, we
first need to account for any correlated noise introduced by the resampling of
individual pixels in the final data cube. Such correlated noise typically results
in an underestimate of the effective noise inside an aperture and, thus in an  overestimate of the real $S/N$ of a source \citep{Gawiser2006ApJS..162....1G,Fumagalli2014MNRAS.444.1282F}. 

To model the noise variation as a function of aperture size, we compute
the effective noise $\sigma_{eff}$ as the standard deviation of fluxes
by considering only the regions free from the continuum detected
sources across the PCWI data cube over a cubic apertures of 4 spectral
pixels ($\sim$ 4~\AA) and a variable aperture size in the spatial
direction. Fig. \ref{fig:noise} shows the ratio between this effective noise
($\sigma_{eff}$) and the error computed by propagating the variance
($\sigma_N$) as a function of apertures size. Ratios above unity indicate that the pipeline noise is underestimated by a factor of $\sim$ 50\% for an aperture of $\sim
3$~arcsec. We account for this effect throughout our analysis.

Focusing on the properties of the emission line next, a single component Gaussian fit to the \lya line gives an intrinsic FWHM (i.e. deconvolved from instrumental effects) of $\approx 131$ \kms\ and a velocity dispersion of $\sigma \approx 56$ \kms.
 The \lya emission is found to be blueshifted from the systemic
 redshift of \zabs$\approx 3.0091$ derived from metal absorption lines (see
 upper three panels of Fig.~\ref{fig:dlaabs}) by $281 \pm 43$ \kms. The flux of
 the \lya line is found to be $f_{\rm Ly\alpha} = \rm (4.9\ \pm 0.9) \times
 10^{-17} erg\ s^{-1}\ cm^{-2}$ detected at 5.3$\sigma$ level which
 reduces to 3.9$\sigma$ level after accounting for the correlated noise (see Fig~\ref{fig:noise}). This corresponds to a \lya  luminosity of \llya $\rm = (3.8 \pm 0.7) \times  10^{42}\ erg\ s^{-1}$ at the DLA redshift. 
  Although the detection significance is marginal from a statistical point of view, the presence of emission feature exactly at the expected location from the SDSS spectrum further strengthen the case for real detection.

In order to identify the location of the galaxy responsible for \lya
emission, we generate a \lya emission map integrated over a velocity
window from $v = -450$ \kms\ to $v = -187$ \kms, comprising the \lya
emission feature.  The left panel of Fig.~\ref{fig:dlamap} shows the
\lya emission map, revealing the location of the DLA host
galaxy. Given the poor spatial resolution of PCWI, we could only place
an upper limit on the extent of the \lya emission to be $< 30$ kpc, at
a surface brightness limit of $\Sigma_{Ly\alpha} > 10^{-17.5}\ \rm
erg\ s^{-1}\ cm^{-2}\ arcsec^{-2}$. The right panel of
Fig.~\ref{fig:dlamap} shows instead the quasar image in the continuum
after collapsing the cube. The overlayed contours mark the \lya
emission map at the flux levels of 0.35, 0.45, 0.55 $\times 10^{-17}
\rm erg\ s^{-1} cm^{-2}$.

It is clear from the figure that the peak flux of the \lya emission
(marked as cross symbol) is off centered from quasar (marked as
diamond).  More quantitatively, we have calculated the separation
between quasar and DLA host galaxy based on light-weighted center,
finding $0.6$~arcsec with a corresponding projected distance of
$5$~kpc. Given that the offset is less than the pixel scale we
consider this as a lower limit. In addition, constrained by the pixel
size we measure an upper limit on impact parameter of 12 kpc (see
also, Fig.\ref{fig:nhivsb}).

\section{Discussion and Conclusion} 

We report the detection of \lya\ emission from the host galaxy of a
DLA with log\nhi\ $ \rm[cm^{-2}] =21.0\pm0.10$ at \zabs$ = 3.0091$
toward the background quasar \qso\ at \zemi= $3.226$. The DLA host is
detected, by selection, at a small impact parameter of $\lesssim 12$ kpc with \lya
luminosity of \llya $\rm = (3.8 \pm 0.7) \times 10^{42}\ erg\ s^{-1}$,
which is typical of the characteristic luminosity (log L$_{\star}
\rm[erg\ s^{-1}]\ =\ 42.66$) of the \lya emitter galaxies at $z \sim 3$ \citep{Herenz2019A&A...621A.107H}.

Given the resonant scattering nature of the \lya line, the emergent
profile is modified and suppressed by many physical effects, e.g. \hi
content, gas geometry and kinematics, and the dust content and
distribution. For instance, in an optically-thick static medium \lya
escapes through successive resonance scattering leading to a
double-humped profile, with the position of the peaks determined by
column density, temperature, and kinematics of the medium
\citep{Neufeld1990ApJ...350..216N,Dijkstra2014PASA...31...40D}. In
addition, scattering through an inflowing (outflowing) medium leads to
an overall blueshift (redshift) of the \lya profile with enhanced blue
(red) peak and suppressed red (blue) peak
\citep[][, see below]{Dijkstra2006ApJ...649...14D}. In a pure static medium, the
expected velocity offset of the \lya emission is $\sim$ 344 \kms\ if
we assume \hi gas temperatures of $10^4$~K \citep[][see their eq.
  21]{Dijkstra2014PASA...31...40D}. For a column density of
$10^{21}~\rm cm^{-2}$, consistent with this DLA, a velocity of
$\approx 300~\rm km~s^{-1}$ is expected, in line with our observations
($\approx 281$~\kms\ with respect to the systemic redshift derived
from metal absorption lines). \par

 However, a static configuration is perhaps unlikely in real systems,
 and the blue offset seen for this DLA host may arise because of an
 inflowing gas geometry
 \citep[e.g.,][]{Rauch2008ApJ...681..856R,Rauch2011MNRAS.418.1115R}.
 \citet{Dijkstra2006ApJ...649...14D} have modelled the spectra and
 surface brightness distributions for the \lya radiation from
 collapsing protogalaxies.  They demonstrate that due to transfer of
 energy from the collapsing gas to the \lya photons, together with a
 reduced escape probability for photons in there wing, causes the blue
 peak to be significantly enhanced, which results in an effective
 blueshift of the \lya line. Furthermore, employing a
 three-dimensional \lya radiative transfer code,
 \citet{Laursen2007ApJ...657L..69L} have investigated the properties
 of young \lya emitting galaxies at high redshift (z $\sim$ 3) from a
 cosmological galaxy formation simulation and found the dominant blue
 peak showing the signature of infalling gas \citep[see
   also,][]{Laursen2009ApJ...704.1640L}.  Such \lya emission profiles
 with prominent blue peaks and suppressed red peaks, with typical
 offset of a few hundred \kms\ as seen here, have been observed in
 several \lya blobs
 \citep{Bower2004MNRAS.351...63B,Wilman2005Natur.436..227W}, LAEs
 \citep{Bunker2003MNRAS.342L..47B} as well as one high confidence DLA
 by \citet{Mackenzie2019MNRAS.tmp.1435M} in VLT-MUSE observations.

Absorption lines from C{\sc~ii}~1334\AA, Fe{\sc~ii}~1608\AA, and
Al{\sc~ii}~1670\AA\ ions are detected in the SDSS spectrum with an
equivalent width of $0.36 \pm 0.04$\AA, $0.25 \pm 0.08$\AA, and
$0.59\pm 0.09$\AA, respectively. Based on the strong correlation seen
between the rest-frame equivalent width of Si{\sc~ii}$\lambda$1526
line and metallicity, we derive an upper limit on the metallicity of
log$ Z/Z_{\odot} < -1.58$ from the observed $3\sigma$ upper limit of
0.34\AA\ on Si{\sc~ii}$\lambda$1526 equivalent width
\citep{Prochaska2008ApJ...672...59P,
  Jorgenson2013MNRAS.435..482J,Neeleman2013ApJ...769...54N}.  In
  addition, we also derive a metallciity of $\simeq -$2.13 based on
  C{\sc~ii}$\lambda$1334 line using the apparent optical depth
  method. This limit also places the DLA just below the average
metallicity of the population at this redshift
\citep{Rafelski2012ApJ...755...89R}, and well below the typical
metallicity of DLAs chosen for targetted searches towards enriched
systems, which typically select DLAs with log $ Z/Z_{\odot} > -1.0$
\cite[see,][]{Krogager2017MNRAS.469.2959K}.  To date, only one
  other DLA host at this low metallicity, namely J$2358+0149$ with log
  Z/Z$\odot$ $\simeq -$1.7 derived based on  Si II absorption, is
detected at small impact parameter of $< 15$~kpc
\citep{Srianand2016MNRAS.460..634S}.

Besides the detection of emission coincident with the DLA position, 
our PCWI search has allowed us to trace the environment of this DLA 
 out to 80 kpc. However, apart from the detection of a fairly bright DLA host, no other galaxies are seen in the field, to a luminosity limit of $5.3 \times 10^{41} \rm erg\ s^{-1}$. 
  This is a somewhat rare occurrence compared to previous studies that have examined the large-scale environment of DLAs. For instance, efforts to directly image the DLA host galaxies show a very small probability of DLAs being associated with bright Lyman-break galaxies at distances $<10-20~\rm kpc$. 
These studies favour instead associations with either faint, possibly  isolated, star-forming galaxies or dwarf galaxies which are  clustered with more massive LBGs
  \citep[see,][]{Fumagalli2015MNRAS.446.3178F}. 
   Moreover, in a recent MUSE
survey of 6 high redshift ($z > 3$) quasar sightlines with 
\hi column density ranging between 20.3 $\le$ log\nhi $ \rm[cm^{-2}] \le 21.15$ 
\citet{Mackenzie2019MNRAS.tmp.1435M} have traced the environment of
DLA host galaxies out to 250 kpc, detecting 5 high-confidence \lya
emitting galaxies associated with three DLAs and 9 lower-significance \lya emission objects in five sightlines.
The MUSE detections are typically found at relatively large impact parameters of $> 50$~kpc, implying that  DLAs generally trace the neutral gas in a wide variety of rich environments, including overdense structures with multiple members.

  In Fig.\ref{fig:nhivsb}, we explore in more detail the known
anticorrelation between the impact parameters versus \nhi. Considering only the high-confidence DLA associations from
the literature and the IFU-based searches probing large scale
environments, a clear trend seems to emerge, although with large
  scatter. Indeed, high \nhi\ systems are observed at preferentially
  small impact parameters, with a Pearson's correlation coefficient of  $-0.544$ and a $p$ value of 0.029 \citep{Zwaan2005MNRAS.364.1467Z,Peroux2011MNRAS.410.2251P,Rubin2015ApJ...808...38R,Krogager2017MNRAS.469.2959K}. It should  however be noted that some of these detections rely on long-slit spectroscopic measurements, for which only small impact parameters are accessible. Larger samples studied with large format IFUs are needed to confirm the significance of this relation.

As a final point, we infer a value for the \emph{in-situ} star
formation rate (SFR, $(\dot{M}_{\rm SF})$) of this DLA host, by
assuming that the \lya photons mainly originate from H~{\sc ii}
regions around massive stars embedded in the DLAs. Assuming case-B
recombination \citep{Osterbrock2006agna.book.....O},
\begin{equation}
L_{\rm Ly\alpha} = 0.68 h\nu_\alpha (1-f_{esc}) N_\gamma \dot{M}_{\rm
  SF},
\label{eq:sfr}
\end{equation}
where $h\nu_\alpha$=10.2~eV [$\rm erg\ s^{-1}$] is the energy of the \lya photons,
$f_{esc}$ is the fraction of ionising photons that escape before
giving rise to ionisation and $N_\gamma$ represents the number of
ionising photons released per baryon of star-formation.

At the redshift of interest of this work, the escape fraction of Lyman
continuum photons is found to vary over a range of $0 \lesssim f_{esc}
\lesssim 0.29$ with an average value of $f_{esc} \sim 0.09$, as
inferred by the Keck Lyman Continuum Spectroscopic Survey of
star-forming galaxies at $z \sim 3$ 
\citep{Steidel2018ApJ...869..123S}. We further assume $N_\gamma =
9870$, corresponding to the average metallicity, i.e. log$Z/Z_{\odot}
= -1.5$, of high redshift DLA absorbers and a Salpeter initial mass
function with $\alpha = 2.35$, as given in \citet[][and references
  therein]{Rahmani2010MNRAS.409L..59R}. The observed \lya luminosity
also depends on the escape fraction ($f_{esc}^{Ly{\alpha}}$) of \lya
photons, and it is related to the emitted \lya luminosity ($\rm
L_{Ly{\alpha}}$) as $ L_{Ly{\alpha}}^{obs} = f_{esc}^{Ly{\alpha}}
L_{Ly{\alpha}}$. The \lya escape fraction increases smoothly and
monotonically out to $z \sim 6$ and strongly depends on the dust
content \citep{Hayes2011ApJ...730....8H}. At the redshift of our
interest the $f_{esc}^{Ly{\alpha}}\approx 5 \pm 3\%$, as is estimated
for the high-redshift ($z \sim 3$) star forming galaxies by
\citet[][see their table
  1]{Hayes2010Natur.464..562H,Hayes2011ApJ...730....8H}.

    Following this method, we infer that the DLA host galaxy is
    forming stars at $\dot{M}_{\rm SF} \approx 21 M_\odot$ yr$^{-1}$
    for an average $f_{esc}\approx 9\%$ and
    $f_{esc}^{Ly{\alpha}}\approx 5\%$. However, this value is highly
    uncertain and ranges between $2 \le \dot{M}_{\rm SF} \le 53$
    [M$_\odot$ yr$^{-1}$] if we account for the large uncertainty in
    the Ly$\alpha$ escape fraction from 70\% to 2\%, respectively
    \citep[see also,][]{Kimm2019MNRAS.486.2215K}.  A similar SFR
      of $\sim 2 M_\odot$ yr$^{-1}$ is also inferred from the star
    formation rate calibration for $H\alpha$ luminosity [i.e.
      $L_{H\alpha}$[erg/s] = 10$^{41.27}$ * SFR $M_\odot$ yr$^{-1}$] from
    \citet{Kennicutt2012ARA&A..50..531K} and the intrinsic
    $Ly\alpha/H\alpha$ ratio of 8-10. This is comparable with the
    typical SFR of Lyman break galaxies at the similar redshifts
    \citep{Kornei2010ApJ...711..693K}.

In conclusion, our search for DLA hosts of high column density (log
$N$ (H{\sc~i}) $\ge 21$) systems, with no metallicity pre-selection
but identified on the basis of likely presence of \lya emission in the SDSS fiber,
appears to be effective in uncovering the gas-galaxy connection in an
interesting region of parameter space, where we expect a direct link
between gas in absorption and star formation in emission
\citep{Rafelski2012ApJ...755...89R,Rafelski2016ApJ...825...87R}. Therefore, future IFU
observations (e.g., PCWI, MUSE, KCWI) of our sample are likely to yield
additional bright DLA host galaxies at small impact parameters, with
which we can start to investigate more systematically both the galaxy
population on large scales, and how neutral gas relates directly to
star formation on smaller scales.

\section*{Acknowledgments}
This work was supported by the National Key R\&D Program of China
(2016YFA0400702, 2016YFA0400703), the National Science Foundation of
China (11473002, 11721303, 11533001) and China Postdoctoral Science
Foundation Grants (2018M630024, 2019T120011). MF acknowledges support
by the Science and Technology Facilities Council [grant number
  ST/P000541/1]. This project has received funding from the European
Research Council (ERC) under the European Union's Horizon 2020
research and innovation programme (grant agreement No 757535). This
project has received funding from the European Research Council (ERC)
under the European Union's Horizon 2020 research and innovation
programme (grant agreement No 757535). This work has been supported by
Fondazione Cariplo, grant No 2018-2329. PN acknowledges support from
the French {\sl Agence Nationale de la Recherche} under grant no
ANR-17-CE31-0011-01. Observations obtained with the Hale Telescope at
Palomar Observatory were obtained as part of an agreement between the
National Astronomical Observatories, Chinese Academy of Sciences, and
the California Institute of Technology. Funding for the Sloan Digital
Sky Survey IV has been provided by the Alfred P. Sloan Foundation, the
U.S. Department of Energy Office of Science, and the Participating
Institutions. SDSS-IV acknowledges support and resources from the
Center for High-Performance Computing at the University of Utah. The
SDSS web site is www.sdss.org. SDSS-IV is managed by the Astrophysical
Research Consortium for the Participating Institutions of the SDSS
Collaboration including the Brazilian Participation Group, the
Carnegie Institution for Science, Carnegie Mellon University, the
Chilean Participation Group, the French Participation Group,
Harvard-Smithsonian Center for Astrophysics, Instituto de
Astrof\'isica de Canarias, The Johns Hopkins University, Kavli
Institute for the Physics and Mathematics of the Universe (IPMU) /
University of Tokyo, Lawrence Berkeley National Laboratory, Leibniz
Institut f\"ur Astrophysik Potsdam (AIP), Max-Planck-Institut f\"ur
Astronomie (MPIA Heidelberg), Max-Planck-Institut f\"ur Astrophysik
(MPA Garching), Max-Planck-Institut f\"ur Extraterrestrische Physik
(MPE), National Astronomical Observatories of China, New Mexico State
University, New York University, University of Notre Dame,
Observat\'ario Nacional / MCTI, The Ohio State University,
Pennsylvania State University, Shanghai Astronomical Observatory,
United Kingdom Participation Group, Universidad Nacional Aut\'onoma de
M\'exico, University of Arizona, University of Colorado Boulder,
University of Oxford, University of Portsmouth, University of Utah,
University of Virginia, University of Washington, University of
Wisconsin, Vanderbilt University, and Yale University.

\bibliographystyle{aasjournal} 
\bibliography{references}
\end{document}